\documentstyle[prb,aps]{revtex}
\newcommand{\be}{\begin{eqnarray}}
\newcommand{\ee}{\end{eqnarray}}
\newcommand{\OP}{\vec{\psi}}
\newcommand{\SG}{\vec{\sigma}}
\newcommand{\M}{\vec{m}}
\newcommand{\U}{\vec{u}}
\newcommand{\F}{{\cal F}}

\begin{document}

\title{Perturbation Expansion in Phase-Ordering Kinetics: II. N-vector 
Model}
\author{Gene F. Mazenko}
\address{The James Franck Institute and the Department of Physics\\
The University of Chicago\\
Chicago, Illinois 60637 }
\date{\today}
\maketitle
%

%
%
\begin{abstract}

The  perturbation theory expansion presented earlier to describe the
phase-ordering kinetics in the case of a nonconserved scalar order
parameter is generalized to the case of the $n$-vector model.  
At lowest order in this expansion, as in the scalar case, one obtains the
theory due to Ohta,  Jasnow and Kawasaki (OJK).  The second-order
corrections for the nonequilibrium exponents are
worked out explicitly in $d$ dimensions and as a function of the number
of components $n$ of the order parameter.
In the formulation developed here the corrections to the OJK 
results are found to go to zero in the large $n$ and $d$ limits.  Indeed, 
the large-$d$ convergence is exponential.

\end{abstract}


\section{Introduction}

A perturbation theory expansion for treating
the scaling
features of phase-ordering kinetics in unstable systems\cite{2} was 
presented earlier\cite{pert}
for the case of a nonconserved scalar order parameter.  In this paper
this method is extended to the case of a vector order parameter with
$n$ components.  It is
found that spin-wave degrees of freedom introduce some new elements
into the theory.  However, as in the scalar case, the Ohta, Jasnow and Kawasaki 
(OJK) theory\cite{OJK,2} emerges as the zeroth-order approximation 
and at this order
the nonequilibrium indices 
$\lambda$ and $\nu$ do not depend on the number of components, $n$, 
of the order parameter.
The second-order  corrections to these exponents are determined
as functions of $d$ and $n$, and
these corrections vanish for both large
$d$ and $n$.

The theory developed in I and here is a two-step process which builds
on earlier work\cite{TUG}.  First one maps the
original problem for the order parameter $\OP$ onto one for an auxiliary 
field $\M$. The field configurations associated with the properly
chosen auxiliary field
are smoother and therefore easier to treat than for the order parameter. The
second step is to treat the nonlinear field theory satisfied by $\M$.
It is found in both scalar and continuous ($n>1$) cases 
that one must construct,  as one constructs a
fixed-point hamiltonian in critical phenomena,
the equation of motion 
satisfied by the auxiliary field in the scaling regime.
A new aspect in the development is the conjecture that there is a general
relation $\nu =2\lambda -d$ connecting the nonequilibrium
indices.  This relation, and
the self-consistent maintenance of the $t^{1/2}$ growth law for this 
problem,
fixes the form of the auxiliary field equation of motion at second
order in the perturbation theory.  An important consequence of this
procedure is that
the second-order contributions to the indices are exponentially small 
in the large-$d$ limit.  
In the continuous case  it is only the longitudinal part
of the auxiliary field equation of motion which must be
determined self-consistently.  The transverse part is unambiguously
determined by consideration of the spin-wave degrees of freedom.  

The nature of the perturbation expansion introduced in I is 
elucidated  further here.  For general $n$ it can be seen that 
one can develop an 
expansion of the nonlinear terms in the auxiliary field equations of motion 
in terms
of a set of vertices labelled by its number of spin labels\cite{special}.
A vertex with $\ell$ labels can be self-consistently taken to be of
 ${\cal O}(\ell /2-1)$.  This, in turn, leads  to the
result that the $\ell^{th}$ order cumulant is also of order 
${\cal O}(\ell /2 -1)$.
We will refer to this expansion as the {\bf vertex~expansion}.  As in 
the scalar
case, this expansion  is well behaved in the lowest orders of
perturbation theory.  We obtain
nontrivial values for the nonequilibrium exponents at second
order in perturbation theory by exponentiating diverging logarithms 
which are driven, for general dimensionality $d$, by internal time integrations.

\section{Overview}

We study here the phase-ordering kinetics generated by the
time-dependent Ginzburg-Landau (TDGL) 
model satisfied by a nonconserved vector order 
parameter $\psi_{i} (\vec{r},t)$, $(i=1,2\ldots ,n)$,
\be
\frac{\partial \psi_{i}}{\partial t}=
-\Gamma \frac{\delta F}{\delta \psi_{i}}  +\eta_{i}
\label{eq:2}
\ee 
where $\Gamma $ is a kinetic coefficient,  $F$ is a Ginzburg-Landau
effective free energy assumed to be of the form
\be
F=\int ~d^{d}r \biggl( \frac{c}{2}\sum_{i=1}^{n}\sum_{\alpha =1}^{d}
(\nabla_{\alpha} \psi_{i})^{2}
+V(\vec{\psi} )\biggr)
\ee
where $c > 0$ and the potential $V$ is assumed to be of the
symmetric 
degenerate wine-bottle form, $V=V(\vec{\psi}^{2})$.
We expect only these general properties
of $V$ will be important in our analysis.
$\vec{\eta} $ is a thermal noise which is related to $\Gamma$ 
by a
fluctuation-dissipation theorem.
We assume that the quench is from a high temperature ($T_{I}>T_{c}$),
where the system is disordered, to zero temperature where the noise 
can be set to zero ($\vec{\eta} =0$).
It is believed\cite{2} that our final results are independent of the 
exact nature of
the initial state, provided it is a disordered state with short-ranged
correlations.

If we rescale length and times we can put our equation of motion in
the dimensionless form
\be
\Lambda (1) \psi_{i} (1)=-V_{i}[\OP (1)]
\equiv -\frac{\partial V[\OP (1)]}{\partial \psi_{i}(1)} ~~~,
\label{eq:eqomot}
\ee
where the diffusion operator 
\be
\Lambda (1)=\frac{\partial}{\partial t_{1}}-\nabla^{2}_{1}
\label{eq:3} 
\ee
is introduced along with the short-hand notation that $1$ denotes
$({\bf r_{1}},t_{1})$.  The standard $\psi^{4}$ form for the potential is
given by
$V=-\frac{1}{2}\vec{\psi}^{2}+\frac{1}{4}(\vec{\psi}^{2})^{2}$.

For late times, following a quench from the disordered to the
ordered phase, the order-parameter dynamics obey scaling\cite{2} governed by
a single growing length, $L(t)$, which is characteristic of the
spacing between defects. In this scaling regime the order-parameter
correlation function has a universal scaling form
\be                                                          
\label{EQ:OPCOR}
C(12) \equiv \langle \vec{\psi} (1)\cdot \vec{\psi} (2) \rangle
= \psi_{0}^{2} \F(x,t_{1}/t_{2})
\nonumber
\ee
where $\psi_{0}$ is the magnitude of the order-parameter in the
ordered phase.
The scaled length $x$ is defined as
$\vec{x} = (\vec{r}_{1}-\vec{r}_{2})/L(T)$ where, for the 
non-conserved order parameter case considered here, the growth law\cite{1d} 
goes as
$L(T) \sim T^{1/2}$ where 
$T=\frac{1}{2}(t_{1}+t_{2})$.
In the case of the autocorrelation function 
$\vec{r}_{1}=\vec{r}_{2}=\vec{r}$ we have\cite{FH}
\be
\langle \vec{\psi} (\vec{r},t_{1})\cdot \vec{\psi} (\vec{r},t_{2}) \rangle
\approx \left(\frac{\sqrt{t_{1}t_{2}}}{T}\right)^{\lambda}
\ee
where $\lambda$ is a nontrivial nonequilibrium exponent which enters
when 
either $t_{1}$ or $t_{2}$ is much larger than the other.
At equal-times the scaling function  $F(x)\equiv\F(x,1)$ is a
nonanalytic function of $x$ for small $|x|$.  This is best reflected
in the generalized form of 
Porod's law\cite{porod,BP} as expressed in
terms of the Fourier transform
${\cal F}(Q)\approx Q^{-(n+d)}$
for large scaled wavenumber $Q$. The large-$x$ behavior of the scaling
function can, with proper 
definition of $x$, be put in the form
$\F (x) \approx \frac{1}{x^{\nu}}e^{-\frac{1}{2}x^{2}}$
where $\nu$ is a nontrivial subdominant exponent
introduced in Ref.(\onlinecite{TUG}).

Values for $\lambda$ and $\nu$ were found in I which suggested that they are
independent.  
In earlier work in Refs.(\onlinecite{NEQ,WM}),
within the approximation developed in Refs.(\onlinecite{TUG}), the two
exponents were related by
\be
\nu =2\lambda -d ~~~.
\label{eq:9}
\ee
We conjecture here 
that
the relation given by Eq.(\ref{eq:9}) holds more generally and the 
perturbation theory
must be constructed to respect this result.  The argument leading to
this conjecture follows from the assumption that in   
the bulk 
scaling regime and away from defects the only available  vector
is $\vec{\psi }(1)$ and the 
potential contribution to the order parameter equation of motion 
must be of the form
\be
V_{i}(\vec{\psi}(1))=-\frac{a}{L^{2}(1)}\psi_{i}(1)
\label{eq:12}
\ee
where $a$ is an unknown constant and, if we are to have scaling,
the coefficient in Eq.(\ref{eq:12}) must be proportional to $L^{-2}$. 
In Refs.(\onlinecite{TUG}) 
and (\onlinecite{NEQ}) it was
found that $a=\pi /2$.  
With these
assumptions one can follow the development in Ref.(\onlinecite{TUG})
to show 
that the exponent $\nu$ can be written in the form 
\be
\nu=d-\frac{a}{\mu} ~~~,
\label{eq:12c}
\ee
where $\mu$ is the eigenvalue which plays a prominent part in the
theory developed in Ref.(\onlinecite{TUG}).  Similarly, generalizing
the analysis in
Ref.(\onlinecite{NEQ})
for two-time correlation functions
in the limit of large time-separations, one can show
that the nonequilibrium exponent $\lambda$ is given by
\be
\lambda =d-\frac{a}{2\mu} ~~~.
\label{eq:14}
\ee
Eliminating $\frac{a}{2\mu}$ between Eqs.(\ref{eq:12c}) and (\ref{eq:14})
we obtain
the scaling relation given by Eq.(\ref{eq:9}).

\section{Auxiliary field Method for the n-vector Model}

Our goal here is to generalize the  perturbation theory expansion approach
developed in I for a scalar order parameter to the case of general $n$.  
We will again express the order parameter as a sum
of two pieces
\be
\vec{\psi}=\vec{\sigma}+ \vec{u}
\label{eq:14a}
\ee
where, as in Ref.\onlinecite{LM92}, $\vec{\sigma}$ is the ordering component
of the order parameter which depends locally on an $n$-component 
field $\vec{m}$:
$\vec{\sigma}=\vec{\sigma}(\vec{m})$.
The field $\vec{u}$ represents the fluctuations about the ordered
configuration.
The functional dependence of $\vec{\sigma}$ on $\vec{m}$ is determined
as a solution of
the Euler-Lagrange equation for 
the associated
stationary defect problem
\be
\sum_{\ell}\frac{\partial ^{2}\sigma_{i} }{\partial m_{\ell}^{2}}
=V_{i}[\sigma (m)] ~~~,
\label{eq:EL}
\ee
where $\vec{m}$ is taken to be the coordinate and the solution must satisfy
the boundary condition $lim_{|m|\rightarrow\infty}\sigma^{2}=\psi_{0}^{2}$.
If we introduce the notation
\be
\sigma_{i;j_{1}j_{2}\ldots j_{\ell}}
\equiv \frac{\partial^{\ell}\sigma_{i}}
{\partial m_{j_{1}}\partial m_{j_{2}}\ldots\partial m_{j_{\ell}}} 
\nonumber
\ee
then Eq.(\ref{eq:EL}) reads
\be
\sum_{\ell}\sigma_{i;\ell \ell }=V_{i}[\sigma [m]] ~~~.
\label{eq:19}
\ee

One can obtain the defect profile analytically for the case
of a general degenerate potential
for the scalar order parameter case.
In the particular case of a
$\psi^{4}$ potential, one
obtains the usual interfacial kink solution
$\sigma [m] = tanh (m/\sqrt{2})$.
For systems with continuous symmetry, $n > 1$, one does not have a 
closed form solution for
the defect profile even for the $\psi^{4}$ potential. 
However one can make some general statements about the form of the profile
in the ordered bulk regime.
For the lowest {\it charge} defects we can write 
the order-parameter profile in the form
$\vec{\sigma}[\vec{m}]=A(m)\hat{m}$,
then the Euler-Lagrange equation reduces to an equation for the
amplitude $A(m)$ given by
\be
A''+\frac{(n-1)}{m}\left(A'-\frac{A}{m}\right)=\frac{\partial V(A)}{\partial A}
~~~.
\label{eq:25}
\ee
Solutions of Eq.(\ref{eq:25}) for large $m$, in the bulk 
away from defect cores, 
in contrast to the scalar case which has an exponential approach
to the ordered value,  show an algebraic
approach to the ordered value:
\be
A=\psi_{0}\left(1-\frac{\xi^{2}}{m^{2}}+\ldots\right) 
\label{eq:38}
\ee
where
$ \xi^{2}\equiv (n-1)/V''(\psi_{0})$.
This solution requires that
the ordered value of the magnitude of the order parameter be given
by the solution to $V'(\psi_{0})=0$,
and, for the solution to be stable, we require $V''(\psi_{0})=q_{0}^{2}> 0 $.
There are some additional general properties we need in the large
$m$ regime. If we write
\be
V_{i}(\vec{\sigma})
=\frac{\partial V(A)}{\partial A}\frac{\partial A}{\partial \sigma_{i}}
=V'(A)\hat{m}_{i} ~~~,
\label{eq:42}
\ee
and use the expansion
\be
V'(A)=V'(\psi_{0})-\psi_{0}\frac{\xi^{2}}{m^{2}}V''(\psi_{0})+\ldots
=-\psi_{0}\frac{(n-1)}{m^{2}}+\ldots ~~~,
\nonumber
\ee
then we have
\be
V_{i}(\vec{\sigma})=-\hat{m}_{i}\psi_{0}\frac{(n-1)}{m^{2}}+\ldots
~~~.
\label{eq:45}
\ee
We also need the second derivative of the potential with respect to
the order parameter.  Taking the derivative of Eq.(\ref{eq:42}) with
respect to $\sigma_{j}$, we obtain
\be
V_{ij}[\vec{m}]=\delta_{ij}\frac{V'(A)}{A}+\sigma_{i}\sigma_{j}\frac{1}{A}
\left(\frac{V'(A)}{A}\right)' 
=P_{ij}[\vec{m}]\frac{V'(A)}{A}+\hat{m}_{i}\hat{m}_{j}V''(A) ~~~,
\nonumber
\ee
where the transverse projection operator is defined by
\be
P_{ij}[\vec{m}]=\delta_{ij}-\hat{m}_{i}\hat{m}_{j}~~~.
\nonumber
\ee
Evaluating $V_{ij}[\vec{m}]$ in the bulk, where we can use 
Eq.(\ref{eq:38}), we obtain
\be 
V_{ij}[\vec{m}]=q_{0}^{2}\hat{m}_{i}\hat{m}_{j}
-P_{ij}\frac{(n-1)}{m^{2}}
~~~.
\label{eq:56}
\ee
where
$ q_{0}^{2}=(n-1)/\xi^{2}$.

Armed with these results, 
we next discuss the equations of motion governing the
fields $\vec{m}$ and $\vec{u}$.  The idea is to separate the
original order-parameter equation of motion into equations for 
$\vec{m}$ and $\vec{u}$
which 
insure that
we do obtain ordering, fluctuations are small and the zeros of
$\M$ reflect the zeros of the order parameter $\OP$.  The condition
that $\SG$ govern the ordering requires that in the  bulk, away from
any defect cores,
$\vec{\sigma}^{2}\rightarrow\psi_{0}^{2}$  and
$\vec{u}^{2}\rightarrow 0$.  
We expect, for both the scalar and continuous cases, that
$\vec{u}$ controls the transition region between the defect core and the
ordered bulk.  

Let us look first at the simpler scaler order parameter case.
If the form given by Eq.(\ref{eq:14a}) is inserted into the 
equation of motion given by Eq.(\ref{eq:3}), we obtain, without 
approximation,
the equation of motion for $u$:
\be
\Lambda (1)u(1)+\sigma_{1}(1)\Lambda (1)m(1)=-V'[\sigma (1)+u(1) ]
+\sigma_{2}(1)(\nabla m(1))^{2} ~~~.
\label{eq:30}
\ee
We assume that the
equation of motion  satisfied by $m$ is  of the form,
\be
\Lambda (1)m(1)=\Xi (1)  ~~~,
\label{eq:30a}
\ee
where $\Xi (1)$ is a functional of $m(1)$ and, if
naive scaling is to hold, we require $\Xi\approx{\cal O}(L^{-1})$.
How one proceeds when one does not have naive scaling will be discussed
elsewhere.
Using Eq.(\ref{eq:30a}) in Eq.(\ref{eq:30}) for $u$ leads to an
equation of motion for the field $u(1)$:
\be
\Lambda (1)u(1)=
-V'[\sigma (1)+u(1) ]+\sigma_{2}(1)(\nabla m(1))^{2}
-\sigma_{1}(1) \Xi (1)  ~~~.
\label{eq:61}
\ee

Our goal then is to show that we can choose $\Xi (1)$ such that
$u$ is small in the bulk and near a defect.  More quantitatively, in the bulk, 
we have $\sigma [m]\rightarrow \psi_{0} sgn [m]$, while 
$lim_{|m|\rightarrow \infty}u [m]=0$. 
To see how this works and to put some
constraints on $\Xi (1)$, we note, if $u$ is small in the bulk,
then we can self-consistently expand
\be
V'[\sigma (1)+u(1) ]=V'[\sigma (1)]
+V''[\sigma (1)]u(1)+\ldots 
=\sigma_{2}(1)+q_{0}^{2}u(1)+\ldots ~~~.
\nonumber
\ee
The equation of motion for $u$ then takes the form in the bulk:
\be
\left[\Lambda (1)+q_{0}^{2}\right]u(1)=
-\sigma_{2}(1)\left[1-(\nabla m(1))^{2}\right]
-\sigma_{1}(1) \Xi (1)  ~~~.
\label{eq:ueq}
\ee
In working in the bulk 
ordered regime, which makes the dominate contribution to the scaling 
properties,  we can estimates $m\approx L$, $\nabla\approx L^{-1}$,
and $\partial /\partial t\approx L^{-2}$.
Notice on the left-hand side of Eq.(\ref{eq:ueq})
that $u$ has acquired a {\it mass}
and, in the long-time long-distance limit, the term  where $u$ is
multiplied by a constant dominates the
derivative terms and $u$ is given by
\be 
q_{0}^{2}u(1)=
-\sigma_{2}(1)\left[1-(\nabla m(1))^{2}\right]
-\sigma_{1}(1) \Xi (1)  ~~~.
\nonumber
\ee
In the limit of large $|m|$ the derivatives of $\sigma $ go
exponentially to zero and the right-hand side of
Eq.(\ref{eq:ueq}) is
exponentially small.  Clearly we can construct a solution for u
which is also exponentially small in the bulk.
We have then on rather general principles that the field
$u$ must vanish rapidly as one moves into the bulk away from
interfaces.  We then have the  rather weak constraints on $\Xi$:

i.  $\Xi$  must be chosen such that $\M$ grows and the field 
$\vec{\sigma}$  orders.

ii.  If the system  satisfies naive scaling then $\Xi$ must go as 
${\cal O}(L^{-1})$ in the bulk.

The form for $\Xi$ in the bulk which fulfills these requirements is given by
\be
\Xi (1)=sgn(m(1))\left( g_{0}(1)+g_{1}(1)
(\nabla m(1))^{2} +\cdots\right)
\nonumber
\ee
where $g_{0}(1)$ and $g_{1}(1)$ must go as $L^{-1}$ for long times.

Let us now look at how this separation process carries over to the case
of a continuous order parameter $n > 1$.
Inserting Eq.(\ref{eq:14a}) into Eq.(\ref{eq:eqomot}) generates an
equation of motion for $\vec{u}(1)$
of the form
\be
\Lambda (1)u_{i}(1)=- V_{i}(\vec{\sigma}(1)+\vec{u}(1))
-\Lambda (1)\sigma_{i}(1)]
=- V_{i}(\vec{\sigma}(1)+\vec{u}(1))
-\sigma_{i;j}\Lambda (1)m_{j}(1)
+\sigma_{i;jk}\nabla_{\alpha} m_{j}\nabla_{\alpha} m_{k} ~~~.
\label{eq:87}
\ee
We  assume that $\vec{m}$ satisfies the nonlinear equation of motion
\be
\Lambda (1)\vec{m}(1)=\vec{\Xi} (1)  ~~~,
\label{eq:88}
\ee
where $\vec{\Xi}$ has the same interpretation as in the scalar case.
Inserting Eq.(\ref{eq:88}) back into Eq.(\ref{eq:87}) gives the basic
equation of motion for $\vec{u}$ :
\be
\Lambda (1)u_{i}(1)
=- V_{i}(\vec{\sigma}(1)+\vec{u}(1))
-\sigma_{i;j}\Xi_{j} (1)
+ \sigma_{i;jk}\nabla_{\alpha} m_{j}\nabla_{\alpha} m_{k}
~~~.
\label{eq:89}
\ee
In the bulk we self-consistently assume that the potential can be expanded
in powers of $\vec{u}$:
\be
 V_{i}(\vec{\sigma}(1)+\vec{u}(1))= V_{i}(\vec{\sigma}(1))
 + V_{ij}(\vec{\sigma}(1))u_{j}(1))+\ldots~~~.
\label{eq:45a}
\ee
The two terms on the right-hand side of Eq.(\ref{eq:45a}) can be 
simplified using
Eqs.(\ref{eq:45}) and (\ref{eq:56}) and lead to the new form for
Eq.(\ref{eq:89}):
\be
\Lambda (1)u_{i}(1)
=\frac{(n-1)}{m^{2}}\hat{m}_{i}-q_{0}^{2}\hat{m}_{i}u_{L}
+\frac{(n-1)}{m^{2}}u_{i}^{T}
-\sigma_{ij}\Xi_{j}(1)
+\sigma_{i;jk}\nabla_{\alpha} m_{j}\nabla_{\alpha} m_{k}
\label{eq:82}
\ee
where we have divided $\vec{u}$ into its longitudinal and transverse parts:
\be
u_{i}=\hat{m}_{i}u_{L}+u_{i}^{T} ~~~,
\ee
where $\hat{m}\cdot\vec{u}^{T}=0$.
Dotting $\hat{m}_{i}$ into Eq.(\ref{eq:82}), we obtain
the equation determining the longitudinal part of $\vec{u}$,
\be
\hat{m}_{i}\Lambda (1)u_{i}(1)
=\frac{(n-1)}{m^{2}}-q_{0}^{2}u_{L}
+\hat{m}_{i}\sigma_{i;jk}\nabla_{\alpha} m_{j}\nabla_{\alpha} m_{k}
-\hat{m}_{i}\sigma_{i;j}\Xi_{j}~~~.
\label{eq:109}
\ee
In the bulk, starting with $\psi_{i}=\psi_{0}\hat{m}_{i}$, we easily obtain
\be
\sigma_{i;j}=\frac{\psi_{0}}{m}P_{ij}[\vec{m}] ~~~,
\nonumber
\ee
$\hat{m}_{i}\sigma_{i;j}\approx {\cal O}(L^{-3})$,
$\hat{m}_{i}\sigma_{i;j}\Xi_{j}\approx {\cal O}(L^{-4})$
and  term proportional to $\hat{m}_{i}\sigma_{i,j}$ can be dropped in 
Eq.(\ref{eq:109}) compared to the terms of ${\cal O}(1/m^{2})$.
Next, since
\be
\sigma_{i;jk}=-\frac{\psi_{0}}{m^{2}}\left[\delta_{ij}\hat{m}_{k}
+\delta_{ik}\hat{m}_{j}+\delta_{jk}\hat{m}_{i}
-3\hat{m}_{i}\hat{m}_{j}\hat{m}_{k}\right)~~~,
\label{eq:82a}
\ee
in the bulk, we have
\be
\hat{m}_{i}\sigma_{i;jk}\nabla_{\alpha} m_{j}\nabla_{\alpha} m_{k}
=-\frac{\psi_{0}}{m^{2}}P_{jk}\nabla_{\alpha} m_{j}\nabla_{\alpha} m_{k}
~~~.
\nonumber
\ee
Using this result back in Eq.(\ref{eq:109}),
the equation of motion in the bulk for
the longitudinal part of the fluctuation field is given then by
\be
\hat{m}_{i}\Lambda (1)u_{i}(1)
=\frac{(n-1)}{m^{2}}-q_{0}^{2}u_{L}
-\frac{\psi_{0}}{m^{2}}P_{jk}\nabla_{\alpha} m_{j}\nabla_{\alpha} m_{k} ~~~.
\label{eq:89b}
\ee
All derivative terms in Eq.(\ref{eq:89b}) acting on $\vec{u}$ can be 
dropped in comparison with the $q_{0}^{2}u_{L}$ term in the scaling 
regime, and, to lowest order in $L^{-1}$, we can express $u_{L}$ explicitly
in terms of $\vec{m}$:
\be
q_{0}^{2}u_{L}=\frac{1}{m^{2}}\left[(n-1)
-P_{jk}\nabla_{\alpha} m_{j}\nabla_{\alpha} m_{k}\right] ~~~.
\label{eq:90} 
\ee
Because of the mass term,  $q_{0}^{2}>0$, we see indeed that 
$u_{L}$ is of order
$L^{-2}$ and no additional constraints are put on $\vec{\Xi}$.

Let us turn next to the transverse part of $\U$.  Multiplying 
Eq.(\ref{eq:82a})
by the transverse projector $P$ we obtain
\be
P_{{ij}}\Lambda (1)u_{j}(1)
=\frac{(n-1)}{m^{2}}u_{i}^{T}
-P_{{ik}}\sigma_{kj}\Xi_{j}(1)
+P_{{i\ell}}\sigma_{\ell;jk}\nabla_{\alpha} m_{j}\nabla_{\alpha} m_{k}
~~~.
\label{eq:113}
\ee
Then in the bulk
\be
P_{{ik}}\sigma_{kj}\Xi_{j}(1)=\frac{\psi_{0}}{m}P_{ij}\Xi_{j}(1)
=\frac{\psi_{0}}{m}\Xi_{i}^{{T}}(1) ~~~,
\label{eq:115}
\ee
and, using Eq.(\ref{eq:82a}),
\be
P_{{i\ell}}\sigma_{\ell;jk}\nabla_{\alpha} m_{j}\nabla_{\alpha} m_{k}
=-2\frac{\psi_{0}}{m^{2}}(\hat{m}_{k}\nabla_{\alpha} m_{k})
P_{ij}\nabla_{\alpha} m_{j} ~~~.
\label{eq:118}
\ee 
Putting Eqs.(\ref{eq:115}) and (\ref{eq:118}) back into Eq.(\ref{eq:113})
gives
\be
P_{ij}\Lambda (1)u_{j}(1)
=\frac{(n-1)}{m^{2}}u_{i}^{T}
-\frac{\psi_{0}}{m}\Xi_{i}^{T}(1)
-2\frac{\psi_{0}}{m^{2}}(\hat{m}_{k}\nabla_{\alpha} m_{k})
P_{ij}\nabla_{\alpha} m_{j} 
\nonumber
\ee
which governs the transverse fluctuations.
For self-consistency $\vec{u}^{T}$ must be small in the
bulk.  This requires us to choose the transverse component of $\vec{\Xi}$
to be given by
\be
-\frac{\psi_{0}}{m}\Xi_{j}^{{T}}(1)
-2\frac{\psi_{0}}{m^{2}}(\hat{m}_{k}\nabla_{\alpha} m_{k})
P_{ij}\nabla_{\alpha} m_{j}=0 ~~~,
\nonumber
\ee
or
\be
\Xi_{i}^{{T}}(1)
=-\frac{2}{m}(\hat{m}_{k}\nabla_{\alpha} m_{k})
P_{ij}\nabla_{\alpha} m_{j}
~~~.
\label{eq:116}
\ee
With this choice, the equation for the transverse fluctuations is given by
\be
P_{ij}\Lambda (1)u_{j}(1)
=\frac{(n-1)}{m^{2}}u_{i}^{T}  ~~~.
\nonumber
\ee
Thus we have that $u_{i}^{T}$ is generated by any coupling back to
$u_{L}\approx {\cal O}(1/L^{2})$ via $P_{ij}\Lambda (1)u_{j}(1)$. We
can estimate 
\be
u_{i}^{T}\approx L^{2}
P_{ij}\Lambda (1)u_{L}(1)\approx {\cal O}(1/L^{2})~~~,
\nonumber
\ee
and generally $\vec{u}\approx {\cal O}(1/L^{2})$.
The requirement that
the bulk part of the transverse fluctuations $\vec{u}^{T}$ be small
fixes the form of $\vec{\Xi}^{T}$  to be given Eq.(\ref{eq:116}).
This form  
does
not depend on any details of the potential, and  can be simplified.
Consider
\be
\frac{2}{m}\hat{m}_{k}\nabla_{\alpha} m_{k}
=\frac{1}{m^{2}}\nabla_{\alpha} m^{2}=\frac{2}{m}\nabla_{\alpha}m
\nonumber
\ee
and
\be
P_{ij}\nabla_{\alpha} m_{j}
=m\frac{\partial \hat{m}_{i}}{\partial m_{j}}\nabla_{\alpha} m_{j}
=m\nabla_{\alpha} \hat{m}_{i} ~~~.
\nonumber
\ee
Inserting these last two results back into Eq.(\ref{eq:116}) gives
\be
\Xi_{i}^{T}(1)
=-\frac{2}{m}(\nabla_{\alpha} m)
m\nabla_{\alpha} \hat{m}_{i}
=-2\nabla_{\alpha} m \nabla_{\alpha} \hat{m}_{i}
~~~.
\label{eq:97}
\ee

The equation of motion satisfied by $\vec{m}$ is given by
Eq.(\ref{eq:88}).
While the transverse part of $\vec{\Xi}$ is given by Eq.(\ref{eq:97}),
the longitudinal part of $\vec{\Xi}$ is constrained only by the
requirement that it scale as $L^{-1}$.  The precise form for $\Xi_{L}$
in the scaling regime
must be determined self-consistently within perturbation theory.
If we look at the building blocks  in the problem we see that 
the quantities which are of ${\cal O}(1)$ are $\hat{m}$ and
$\nabla_{\alpha}m_{i}$.  Thus one sees that the structure of
the longitudinal part of the equation of motion
in the bulk scaling regime 
can be assumed to be of the general form:
\be
\Xi_{i}^{L}(1)=
\hat{m}_{i}(1)\Biggl[ g_{0}(1)
+g_{1}(1)\left(\nabla_{\alpha} m_{j}(1)\right)^{2} 
+g_{ijk\ell}^{(2)}(1)
\nabla_{\alpha} m_{i}(1)\nabla_{\alpha} m_{j}(1)
\nabla_{\beta} m_{k}(1)\nabla_{\beta} m_{\ell}(1)+\cdots\Biggr]
~~~.
\nonumber
\ee
Clearly in the long-time limit we require the $g$'s be of 
${\cal O}\left(L^{-1}\right)$
and, as we shall see, that we can self-consistently construct the $g_{p}$'s
if we assume that $g_{p}={\cal O}(p)$ in the vertex expansion.
Our final results at second order in our expansion will depend on $g_{0}$ and 
$g_{1}$.

The assumption we make here is that the higher-order terms 
proportional to $g^{(\ell)}$, for $\ell > 0$, contribute in a non-trivial
way starting at order
$\ell +2$.  Thus $g^{(2)}$, due to various contractions, acts at
second order only to renormalize $g_{0}$.

\section{Field Theory for Auxiliary Field}

The equation of motion satisfied by the ordering field
$\M$ including terms which contribute up to second order
 is given in the bulk  by 
\be
\Lambda (1)m_{i}(1)=g_{0} (t_{1}) \hat{m}_{i}(1)) +\tilde{\Xi}_{i}(1)
\label{eq:50}
\ee
where
$\tilde{\Xi}_{i}(1) =\tilde{\Xi}_{i}^{L}(1)+\Xi_{i}^{T}(1)$, and
\be
\tilde{\Xi}_{i}^{L}(1)
=g_{1}(t_{1})\hat{m}_{i}(1)\left(\nabla_{\alpha} m_{j}(1)\right)^{2}
\nonumber
\ee
\be
\Xi_{i}^{T}(1)
=-2\nabla_{\alpha} m(1) \nabla_{\alpha} \hat{m}_{i}(1)
~~~.
\nonumber
\ee
The functions $g_{0}(t_{1})$ and $g_{1}(t_{1})$ are determined within
perturbation theory.
Our  analysis will follow  the standard 
Martin-Siggia-Rose\cite{MSR} method in its functional integral form 
as developed by 
DeDominicis and Peliti\cite{DP}  and presented in detail in I.
In the MSR method the field 
theoretical development requires a doubling of fields to include the 
response field $\vec{M}$.   
As in I, we introduce a field $\vec{h}(1)$ conjugate to $\vec{m}(1)$
and a field $\vec{H}(1)$ conjugate to $\vec{M}(1)$.

Following closely the formal development in I, we find that
the fundamental equation satisfied by the average of the field
$\vec{m}$, in the presense
of sources, is given by
\be
i\left[\Lambda (1) 
<m(1)>_{h}-Q(1)\right] 
=-\int d2 ~\Pi_{0} (12)<M(2)>_{h}+H(1) 
\label{eq:77}
\ee
where the vector labels are suppressed,
\be
\Pi_{0}^{ij} (12) \equiv \delta (t_{1}-t_{0})\delta (t_{1}-t_{2})
g(\vec{r}_{1}-\vec{r}_{2})\delta_{ij} ~~~.
\nonumber
\ee
It is assumed here that
the field $\vec{m}(1)$, at the initial time $t_{0}$ has gaussian 
statistics with variance
\be
<m_{0}^{i}(\vec{r}_{1})m_{0}^{j}(\vec{r}_{2})>=
\delta_{ij}g(\vec{r}_{1}-\vec{r}_{2})
~~~.
\nonumber
\ee
The nonlinear vertices in Eq.(\ref{eq:77}) are given by
\be
Q_{i}(1)=<\Xi_{i}(1)>\equiv Q_{i}^{D}(1)+Q_{i}^{L}(1)+Q_{i}^{T}(1)
\nonumber
\ee
with
\be
Q_{i}^{D}(1)=<\Xi_{i}^{D}(1)>=g_{0}(1)<\hat{m}_{i}(1)>
\nonumber
\ee
\be
Q_{i}^{L}(1)=<\Xi_{i}^{L}(1)>
=g_{1}(1)<\hat{m}_{i}(1)\left(\nabla m_{j}(1)\right)^{2}>
\nonumber
\ee
\be
Q_{i}^{T}(1)=<\Xi_{i}^{T}(1)>
=(-2)<\nabla_{\alpha} m(1)\nabla_{\alpha}\hat{m}_{i}(1)> ~~~.
\nonumber
\ee
The fundamental equation satisfied by the average of the MSR response field
is given by
\be
-i\left[\tilde{\Lambda}(1)<M(1)>_{h}
+\tilde{Q}(1)\right]=h(1)
\label{eq:73}
\ee
where, we define
\be
\tilde{\Lambda}(1)=\frac{\partial}{\partial t_{1}}+\nabla^{2}_{1}
~~~,
\nonumber
\ee
and the nonlinear contributions are given by
\be
\tilde{Q}_{i}(1)=<\bar{\Xi}_{i}(1)>
=\tilde{Q}_{i}^{D}(1)+\tilde{Q}_{i}^{L}(1)+\tilde{Q}_{i}^{T}(1)
\nonumber
\ee
with
\be
\tilde{Q}_{i}^{D}(1)=
<\bar{\Xi}^{D}_{i}(1)>=g_{0}(1) <M_{j}(1)\rho_{ij}(1)>
\nonumber
\ee
\be
\tilde{Q}_{i}^{L}(1)=<\bar{\Xi}^{L}_{i}(1)>
=< M_{j}(1)g_{1}(1)\rho_{ij} (1)
\left(\nabla_{\alpha}m_{k}(1)\right)^{2}>
-<\nabla_{\alpha}\left[ M_{j}(1)g_{1}(1)\hat{m}_{j} (1)
2\nabla_{\alpha}m_{i}(1)\right]>
\nonumber
\ee
and
\be
\tilde{Q}_{i}^{T}(1)=
<\bar{\Xi}^{T}_{i}(1)>
=<2\hat{m}_{i}(1)\nabla_{\alpha}
\left[ M_{j}(1)\nabla_{\alpha}\hat{m}_{j}(1)\right]
+2\rho_{ij}(1)\nabla_{\alpha}\left[M_{j}(1)\nabla_{\alpha}m(1)\right]>
~~~,
\nonumber
\ee
and
\be
\rho_{ij}(1)=\frac{1}{m}\left(\delta_{ij}-\hat{m}_{i}\hat{m}_{j}\right)
~~~.
\nonumber
\ee

All correlation functions of interest can be
generated as functional derivatives of $<m(1)>_{h}$ or $<M(1)>_{h}$
with respect to $h(1)$ and $H(1)$.
In the limit in which  the source fields vanish, each term in the
two fundamental equations vanish.  Therefore it is derivatives 
with respect to the external sources of these
equations which are of interest.   
Let us introduce the notation that
$G_{A_{1},A_{2},....,A_{n}}(12...n)$ is the $n^{th}$ order cumulant
for the set of fields $\{A_{1},A_{2},....,A_{n}\}$ where field
$A_{1}$ has argument $(1)$, field $A_{2}$ has argument $(2)$, et cetera.
This notation is needed when have cumlants with $m$ and $M$ mixed.
As an example
\be
G_{Mmmm}(1234)=\frac{\delta^{3}<m(4)>_{h}}{\delta H(1)
\delta h(2)\delta h(3)} ~~~.
\label{eq:120}
\ee
As a short hand for cumulants involving only $m$ fields we write
\be
G^{(n)}(12\cdots n)=\frac{\delta^{n-1}}{\delta h(n)\delta h(n-1)
\cdots \delta h(2)}<m(1)>_{h} ~~~.
\label{eq:121}
\ee
The hierarchy of equations connecting these cumulants is given by taking
functional derivatives of the fundamental equations given by
Eqs.(\ref{eq:77}) and (\ref{eq:73}).

The equations governing the $n^{th}$ order cumulants are given by
\be
-i\left[\tilde{\Lambda}(1)G_{Mm...m}(12...n)
+\tilde{Q}_{n}(12...n)\right]=0
\label{eq:125}
\ee
and
\be  
i\left[\Lambda (1)G^{(n)}(12...n)-Q_{n}(12...n)\right]  
=-\int 
d\bar{1} ~\Pi_{0} (1\bar{1})G_{Mm...m}(\bar{1}2...n) ~~~,
\label{eq:92}
\ee
where the $Q's$ are defined by
\be
\tilde{Q}_{n}(12...n)=
\frac{\delta^{n-1}}{\delta h(n)\delta h(n-1)\cdots\delta h(2)}
\tilde{Q}(1)
\label{eq:88a}
\ee
and
\be
Q_{n}(12...n)=
\frac{\delta^{n-1}}{\delta h(n)\delta h(n-1)\cdots\delta h(2)}Q (1)~~~.
\label{eq:89a}
\ee
With this notation the equations determining the two-point functions can be
written as
\be
-i\left[\tilde{\Lambda}(1)G_{Mm}(12)
+\tilde{Q}_{2}(12)\right]=\delta (12)
\label{eq:133}
\ee
\be  
i\left[\Lambda (1)G(12)-Q_{2}(12)\right]  
=-\int 
d\bar{1} ~\Pi_{0} (1\bar{1})G_{Mm}(\bar{1}2)
\label{eq:134}
\ee

The point now is to show that there is a consistent perturbation expansion
for this theory.  
To get started we need to express $\hat{Q}_{i}(1)$ and $Q_{i}(1)$
in terms of a fundamental set of vertices which can be  written 
in terms of the
singlet probability distribution
\be
P_{h}(\vec{x},1)=<\delta (\vec{x}-\vec{m}(1))>_{h}
~~~.
\nonumber
\ee
After a great deal of rearrangement one can show that the 
nonlinear vertices, the $Q$'s can
be put in the form
\be
Q_{i}^{D}(1)=g_{0}(1)U_{i}(1)
\label{eq:136}
\ee
\be
Q_{i}^{L}(1)=\sum_{j}\int d\bar{2} d\bar{3}~ 
g_{1}(1)w(1\bar{2}\bar{3})O_{jj}(\bar{2}\bar{3})U_{i}(1)
\nonumber
\ee
\be
Q_{i}^{T}(1)=\sum_{jk}\int d\bar{2} d\bar{3}~
w(1\bar{2}\bar{3})O_{jk}(\bar{2}\bar{3})U_{jk,i}(1)
~~~.
\nonumber
\ee
\be
\tilde{Q}_{i}^{D}(1)=- \sum_{j}\int d\bar{2} d\bar{3}~
g_{0}(1)O_{H_{j}}(1)U_{i,j}(1) ~~~.
\nonumber
\ee
\be
\tilde{Q}_{i}^{L,1}(1)=- \sum_{jk}\int d\bar{2} d\bar{3}~
g_{1}(1)w(1\bar{2}\bar{3})O_{kk}(\bar{2}\bar{3})
O_{M_{j}}(1)U_{i,j}(1)
\nonumber
\ee
\be
\tilde{Q}_{i}^{L,2}(1)=-\sum_{j}\int d\bar{2} d\bar{3}~
2g_{1}(1)\tilde{w}(1\bar{2}\bar{3})
O_{i}(\bar{2})O_{M_{j}}(\bar{3})U_{j}(\bar{3})
\nonumber
\ee
\be
\tilde{Q}_{i}^{T,1}(1)=-\sum_{js\ell}\int d\bar{2} d\bar{3}~
w(1\bar{2}\bar{3})O_{M_{j}}(1)
O_{s\ell}(\bar{2}\bar{3}){\cal Q}_{s\ell ,ij}(1)
\nonumber
\ee
\be
\tilde{Q}_{i}^{T,2}(1)=-\sum_{js\ell}\int d\bar{2} d\bar{3}~
2\tilde{w}(1\bar{2}\bar{3})O_{M_{j}}(\bar{3})
O_{\ell}(\bar{2})U_{i\ell ,j}(1) 
\label{eq:142}
\ee
where we have introduced the operators
\be
O_{j}(2)=\frac{\delta}{\delta h_{j}(2)}+G^{(1)}_{j}(2)
~~~,
\nonumber
\ee
and
\be
O_{jk}(23)=\frac{\delta^{2}}{\delta h_{j}(2)\delta h_{k}(3)}
+G^{(2)}_{jk}(23)+G^{(1)}_{j}(2)\frac{\delta}{\delta h_{k}(3)}
\nonumber
\ee
\be
+G^{(1)}_{k}(3)\frac{\delta}{\delta h_{j}(2)}+G^{(1)}_{j}(2)G^{(1)}_{k}(3)
~~~.
\nonumber
\ee
We have also introduced
the three-point vertices
\be
w(123)=\sum_{\alpha =1}^{d} \nabla_{\alpha}^{(1)}\delta (12)
\nabla_{\alpha}^{(1)}\delta (13) 
\nonumber
\ee
and
\be
\tilde{w}(123)=\nabla_{\alpha}¥^{(1)}¥\left[\delta (13)
\nabla_{\alpha}¥^{(1)}\delta (12)\right]
~~~.
\nonumber
\ee
Each term in these expressions for the $Q$'s can be expressed in terms of
the set of nonlinear vertices which are integral moments of 
$P_{h}[\vec{x}]$:
\be
U_{ijk\ldots,\ell mn\ldots}(1)=
\int ~d^{n}x ~\hat{x}_{i}\hat{x}_{j}\hat{x}_{k}\ldots
\nabla_{x}^{\ell}\nabla_{x}^{m}\nabla_{x}^{n}
\ldots P_{h}[\vec{x},1]~~~.
\label{eq:147}
\ee
We have also defined
\be
{\cal Q}_{s\ell ,ij}(1)=U_{s\ell ,i j}(1)-U_{is,\ell j}(1)-
U_{i\ell ,s j}(1) ~~~.
\nonumber
\ee

\section{Perturbation Theory Expansion}

All of the cumulants involving
the field $\vec{m}$ can, in principle, be obtained from 
Eqs.(\ref{eq:77}) and (\ref{eq:73}) by taking functional derivatives.
This then requires that we work
out the functional derivatives of $Q_{n}$ and $\tilde{Q}_{n}$ which
are defined by Eqs.(\ref{eq:88a}) and (\ref{eq:89a}).  These
objects are functional derivatives of $Q_{1}$ and $\tilde{Q}_{1}$
which are proportional to a few of the
$U_{ijk\ldots,\ell mn\ldots}(1)$ and functional derivatives of
these quantities.  From this discussion it should be clear that
all of the
$Q_{n}$ and $\hat{Q}_{n}$ can be written as a product of cumulants
multiplying vertices given by
\be
U_{ijk\ldots,\ell mn\ldots, stu\ldots}(1;234\ldots)=
\frac{\delta}{\delta h_{s}(2)}\frac{\delta}{\delta h_{t}(3)}
\frac{\delta}{\delta h_{u}(4)}\ldots U_{ijk\ldots,\ell mn\ldots}(1)
~~~.
\nonumber
\ee
The point we want to establish is that if $U$ has $p$ vector labels, 
then, at lowest order,
we can take $U$ to be of ${\cal O}(p/2-1)$, plus higher order terms.

The perturbation theory expansion for the 
$U_{ijk\ldots,\ell mn\ldots}(1)$ follows from the expansion properties of the
singlet-distribution function.
The perturbation theory expansion for $P_{h}(\vec{x},1)$ is straightforward.  
Using the
integral representation for the $\delta$-function, we have
\be
P_{h}(\vec{x},1)= \int\frac{d^{n}k}{(2\pi)^{n}}e^{-i\vec{k}\cdot\vec{x}}
<e^{{\cal H}(1)}>_{h}
\nonumber
\ee
where
${\cal H}(1)\equiv i\vec{k}\cdot\vec{m}(1)$.
The average of the exponential is precisely of the form which can be
rewritten in terms of cumulants:
\be
\Phi (\vec{k},1)\equiv <e^{{\cal H}(1)}>_{h}
=exp\left[\sum_{s=1}^{\infty}\frac{1}{s!}G_{{\cal H}}^{(s)}(1)\right]
\nonumber
\ee
where
$G_{{\cal H}}^{(s)}(1)$ is the $s^{th}$ order cumulant for the field 
${\cal H}(1)$.
Since ${\cal H}(1)$ is proportional to $\vec{m}(1)$ these are, up to factors
of $i\vec{k}$ to the $s^{th}$ power, just the cumulants for the $m$ 
field:
\be
G_{{\cal H}}^{(1)}(1)=i\sum_{\alpha_{1}} k_{\alpha_{1}}G^{(1)}_{\alpha_{1}}(1)
\nonumber
\ee
\be 
G_{{\cal H}}^{(2)}(1)=(i)^{2}\sum_{\alpha_{1}\alpha_{2}}
k_{\alpha_{1}}k_{\alpha_{2}}G^{(2)}_{\alpha_{1}\alpha_{2}}(11)
\nonumber
\ee
\be  
G_{{\cal H}}^{(3)}(1)=(i)^{3}\sum_{\alpha_{1}\alpha_{2}\alpha_{3}}
k_{\alpha_{1}}k_{\alpha_{2}}k_{\alpha_{3}}
G^{(3)}_{\alpha_{1}\alpha_{2}\alpha_{3}}(111)
\nonumber
\ee
and so on.  We can therefore write
\be
\Phi (\vec{k},1)
=exp\left[\sum_{s=1}^{\infty}\frac{(i)^{s}}{s!}
k_{\alpha_{1}}k_{\alpha_{2}}\cdots k_{\alpha_{s}}
G^{(s)}_{\alpha_{1}\alpha_{2}\alpha_{3}\ldots\alpha_{s}}(11...1)\right] ~~~.
\nonumber
\ee

We will assume, as we will show self-consistently, that $p^{th}$ order 
cumulants are of order $\frac{p}{2}-1$ in the vertex expansion. 
Expanding and
keeping
terms up to the 4-point cumulant, we obtain
\be
P_{h}(\vec{x},1)=\left[1-\sum_{\alpha_{1}\alpha_{2}\alpha_{3}}
\frac{1}{3!}G^{(3)}_{\alpha_{1}\alpha_{2}\alpha_{3}}(111)
\nabla_{x}^{\alpha_{1}}\nabla_{x}^{\alpha_{2}}\nabla_{x}^{\alpha_{3}}
+\cdots\right] P_{h}^{(0)}(\vec{x},1)
\label{eq:151}
\ee
where
\be
P_{h}^{(0)}(\vec{x},1)=\int\frac{d^{n}k}{(2\pi)^{n}} 
\Phi_{h}^{(0)}(\vec{k},1)e^{-i\vec{k}\cdot\vec{x}}
\label{eq:159}
\ee
and
\be
\Phi_{h}^{(0)}(\vec{k},1)=e^{i\sum_{\alpha_{1}}k_{\alpha_{1}}
G^{(1)}_{\alpha_{1}}(1)}
e^{-\frac{1}{2}\sum_{\alpha_{1}\alpha_{2}}k_{\alpha_{1}}k_{\alpha_{2}}
G^{(2)}_{\alpha_{1}\alpha_{2}}(11)} ~~~.
\nonumber
\ee
We can define the  lowest order set of vertices
\be
U_{ijk\ldots,\ell mn\ldots}^{(0)}(1)=
\int ~d^{n}x ~\hat{x}_{i}\hat{x}_{j}\hat{x}_{k}\ldots
\nabla_{x}^{\ell}\nabla_{x}^{m}\nabla_{x}^{n}
\ldots P_{h}^{(0)}[\vec{x},1]
\nonumber
\ee
\be
=\int ~d^{n}x ~\hat{x}_{i}\hat{x}_{j}\hat{x}_{k}\ldots 
\nabla_{x}^{\ell}\nabla_{x}^{m}\nabla_{x}^{n}\ldots 
\int\frac{d^{n}k}{(2\pi)^{n}}
\Phi_{h}^{(0)}(\vec{k},1)e^{-i\vec{k}\cdot\vec{x}}
\nonumber
\ee
\be
=\int ~d^{n}x ~\hat{x}_{i}\hat{x}_{j}\hat{x}_{k}\ldots 
\int\frac{d^{n}k}{(2\pi)^{n}}
(-ik_{\ell})(-ik_{m})(-ik_{n})\ldots 
\int\frac{d^{n}k}{(2\pi)^{n}}
\Phi_{h}^{(0)}(\vec{k},1)e^{-i\vec{k}\cdot\vec{x}} ~~~.
\label{eq:159a}
\ee
It should be clear, after inserting Eq.(\ref{eq:151}) back into
Eq.(\ref{eq:147}), that
\be
U_{ijk\ldots,\ell mn\ldots}(1)=U_{ijk\ldots,\ell mn\ldots}^{(0)}(1)
+\sum_{\alpha_{1}\alpha_{2}\alpha_{3}\alpha_{4}}
U_{ijk\ldots,\ell mn\ldots\alpha_{1}\alpha_{2}\alpha_{3}\alpha_{4}}^{(0)}(1)
\frac{1}{4!}G^{(4)}_{\alpha_{1}\alpha_{2}\alpha_{3}\alpha_{4}}(1111)
+\ldots ~~~.
\label{eq:160}
\ee
It is clear that  if the term on the first line of Eq.(\ref{eq:160})
 is of ${\cal O}(p)$,
then the term on the second line is of ${\cal O}(p+3)$

For these ideas to be self-consistent then the functional
derivatives of $U^{(0)}{\ldots}$ must be higher order.  The reason this
works is because factors of the one-point cumulant $G_{i}(1)$ enters
in the exponential appearing in  $\Phi_{h}^{(0)}(\vec{k},1)$ multiplied
by a factor of $\vec{k}$.  Thus functional derivatives either bring down
factors of $\vec{k}$  from the exponential or increase the order of
cumulants which do not involve the one-point cumulant.
To see how this works consider  the set of derivatives
\be
U_{\ldots\alpha_{3}\alpha_{4}}(1;34)=
\frac{\delta^{2}}{\delta h_{\alpha_{3}}(3)\delta h_{\alpha_{4}}(4)}
U^{(0)}_{\ldots}(1)~~~.
\nonumber
\ee
The functional derivatives then act on 
$\Phi_{h}^{(0)}(\vec{k},1)$.  It is then easy enough to work out,
using Eqs.(\ref{eq:159a})  and (\ref{eq:159}), that 
\be
\frac{\delta^{2}}{\delta h_{\alpha_{3}}(3)\delta h_{\alpha_{4}}(4)}
\Phi_{h}^{(0)}(\vec{k},1)|_{h=0}
=\sum_{\alpha_{1}\alpha_{2}}k_{\alpha_{1}}k_{\alpha_{2}}
\left[G^{(2)}_{\alpha_{1}\alpha_{3}}(13)G^{(2)}_{\alpha_{2}\alpha_{4}}(14)
-\frac{1}{2}G^{(4)}_{\alpha_{1}\alpha_{2}\alpha_{3}\alpha_{4}}(1134)\right]
\Phi_{0}^{(0)}(\vec{k},1)
~~~.
\nonumber
\ee
We then have
\be
U_{\ldots\alpha_{3}\alpha_{4}}(1;34)
=\sum_{\alpha_{1}\alpha_{2}}U^{(0)}_{\ldots\alpha_{1}\alpha_{2}}(1)
\left[G^{(2)}_{\alpha_{1}\alpha_{3}}(13)G^{(2)}_{\alpha_{2}\alpha_{4}}(14)
-\frac{1}{2}G^{(4)}_{\alpha_{1}\alpha_{2}\alpha_{3}\alpha_{4}}(1134)\right]
\nonumber
\ee
and if $U^{(0)}_{\ldots}(1)$ is of ${\cal O}(p)$ then
$U^{(0)}_{\ldots\alpha_{1}\alpha_{2}}$ is of ${\cal O}(p+1)$,
$G^{(4)}_{\alpha_{1}\alpha_{2}\alpha_{3}\alpha_{4}}$ is of ${\cal O}(1)$, and
$U_{\ldots\alpha_{3}\alpha_{4}}(1;34)$ is of ${\cal O}(p+1)$ plus higher 
order terms.

Notice then that all derivatives of $U$ and higher order terms 
can be expressed as products of the vertices $U^{(0)}$ with legs
connected by cumulants.  Any contribution  to the nonlinear vertices
$Q_{n}$ and $\tilde{Q}_{n}$ can therefore be ordered summing up the
contributions from each vertex and cumulant where a vertex with $2s+2$ 
legs makes a contribution of
${\cal O}(s)$ and a cumulant $G^{(2p+2)}$ makes a contribution
 of ${\cal O}(p)$.

For our purposes here, we only need two sets of the bare vertices $U^{(0)}$:
\be
{\cal V}_{s}(\alpha_{1};\alpha_{2},\ldots\alpha_{2s+2};1)
=\int ~d^{n}x ~\hat{x}_{\alpha_{1}} 
\int\frac{d^{n}k}{(2\pi)^{n}}
(i)k_{\alpha_{2}}k_{\alpha_{3}}\ldots k_{\alpha_{2s+2}}
\Phi_{h}^{(0)}(\vec{k},1)e^{-i\vec{k}\cdot\vec{x}}
\nonumber
\ee
and
\be
\tilde{{\cal V}}_{s}(\alpha_{1},\alpha_{2};\ldots\alpha_{2s+2};1)
=\int ~d^{n}x ~\hat{x}_{\alpha_{1}} \hat{x}_{\alpha_{2}}
\int\frac{d^{n}k}{(2\pi)^{n}}
k_{\alpha_{3}}k_{\alpha_{4}}\ldots k_{\alpha_{2s+2}}
\Phi_{h}^{(0)}(\vec{k},1)e^{-i\vec{k}\cdot\vec{x}}
~~~.
\nonumber
\ee
It can be seen the ${\cal V}_{s}(\alpha_{1};\alpha_{2}\ldots\alpha_{2s+2};1)$
is the natural generalization of the vertex $\phi_{s}(1)$ given by Eq.(92) 
in
I.  As in I, it is straightforward to work out these vertices explicitly by
doing the integrations.  The quantities we will need below are given by:
\be
{\cal V}_{0}(\alpha_{1};\alpha_{2};1)=\delta_{\alpha_{1},\alpha_{2}}
\sqrt{\frac{2}{S(1)}}\frac{\Gamma (\frac{n+2}{2})}{n\Gamma (\frac{n}{2})}
~~~,
\nonumber
\ee
\be
{\cal V}_{1}(\alpha_{1};\alpha_{2},\alpha_{3},\alpha_{4};1)
=I_{\alpha_{1},\alpha_{2},\alpha_{3},\alpha_{4}}
\sqrt{\frac{2}{S^{3}(1)}}
\frac{\Gamma (\frac{n+1}{2})}{n(n+2)\Gamma (\frac{n}{2})}
~~~,
\nonumber
\ee
\be
\tilde{{\cal V}}_{0}(\alpha_{1},\alpha_{2};1)=\delta_{\alpha_{1},\alpha_{2}}
\frac{1}{n}
~~~,
\nonumber
\ee
and
\be
\tilde{{\cal V}}_{1}(\alpha_{1},\alpha_{2};\alpha_{3},\alpha_{4};1)
=\frac{1}{S(1)}\left[\frac{1}{n}\delta_{\alpha_{1},\alpha_{2}}
\delta_{\alpha_{1},\alpha_{2}}-\frac{1}{n+2}
I_{\alpha_{1},\alpha_{2},\alpha_{3},\alpha_{4}}\right]
~~~.
\nonumber
\ee
W have  also introduced
\be
S(1)\equiv G^{(2)}_{\alpha\alpha}(11)=<m_{\alpha}^{2}(1)> 
~~~,
\nonumber
\ee
with no summation over $\alpha$, 
and the symmetric tensor
\be
I_{\alpha_{1},\alpha_{2},\alpha_{3},\alpha_{4}}=
\delta_{\alpha_{1},\alpha_{2}}\delta_{\alpha_{3},\alpha_{4}}
+\delta_{\alpha_{1},\alpha_{3}}\delta_{\alpha_{2},\alpha_{4}}
+\delta_{\alpha_{1},\alpha_{4}}\delta_{\alpha_{2},\alpha_{3}}
~~~.
\nonumber
\ee

\subsection{Two-Point nonlinear vertices to second order}

Armed with these results we can return to the evaluation of the 
$Q_{n}$ and $\tilde{Q}_{n}$ derived from $Q_{1}$ and $\tilde{Q}_{1}$
which are given by Eqs.(\ref{eq:136})-(\ref{eq:142}).  We summarize
first the results for the
two-point quantities which enter Eqs.(\ref{eq:133}) and(\ref{eq:134}).
We find after a significant amount of work that 
\be
Q_{2}(12)=\Omega (1\bar{1})G(\bar{1}2)
+\tilde{S}(1\bar{2}\bar{3}\bar{4})G(\bar{2}\bar{3}\bar{4}2)
\label{eq:169}
\ee
\be 
\tilde{Q}_{2}(12)=\Omega (1\bar{1})G_{Mm}(\bar{1}2) 
+\tilde{S}_{1}(1\bar{2}\bar{3}\bar{4})G_{Mmmm}(\bar{2}\bar{3}\bar{4}2) 
\label{eq:170}
\ee
where integration and summation over repeated barred indices is
implied.  The various auxiliary quantities are defined by:
\be
\Omega (12)=\left(g_{0}(1)+g_{1}(1)S^{(2)}(1)\right)
{\cal V}_{0}(\alpha_{1};\alpha_{2};1)\delta (12)
\label{eq:172}
\ee
\be
\tilde{S}(1234)=\frac{1}{3!}\tilde{V}_{0}(1234)
+\tilde{V}_{L}(1234)+\tilde{V}_{1}(1234)
\label{eq:173}
\ee
\be 
\tilde{S}_{1}(1234)=\frac{1}{2}\tilde{V}_{0}(1234)
+\tilde{V}_{L}(1234)+\tilde{V}_{2}(1234)
+\tilde{V}_{3}(1432)+\tilde{V}_{L}^{T}(1432)
\label{eq:174}
\ee
\be
\tilde{V}_{0}(1234)=-g_{0}(1)
{\cal V}_{1}(\alpha_{1};\alpha_{2},\alpha_{3},\alpha_{4};1)
\delta (12)\delta (13)\delta (14)
\nonumber
\ee
\be
\tilde{V}_{L}(1234)=g_{1}(1){\cal V}_{0}(\alpha_{1};\alpha_{2};1)
\delta_{\alpha_{3}\alpha_{4}}W(134)
\delta (12)
\nonumber
\ee
\be
\tilde{V}_{1}(1234)=
\tilde{{\cal V}}_{1}(\alpha_{3},\alpha_{4};\alpha_{1},\alpha_{2};1)
\delta_{\alpha_{3}\alpha_{4}}W(134) 
\delta (12)
\nonumber
\ee
\be
\tilde{V}_{2}(1234)=-W(134)\delta (12){\cal Q}^{(0)}_{\alpha_{3},\alpha_{4},
\alpha_{1},\alpha_{2}}(1)
\nonumber
\ee
\be
\tilde{V}_{3}(1234)=-2
\tilde{{\cal V}}_{1}(\alpha_{1},\alpha_{3};\alpha_{4},\alpha_{2};1)
\tilde{W}(134)\delta (12)
\nonumber
\ee
\be
\tilde{V}_{L}^{T}(1234)=-2g_{1}(4){\cal V}_{0}(\alpha_{4};\alpha_{2};4)
\delta_{\alpha_{1},\alpha_{3}}\tilde{W}(134)\delta (24)
\nonumber
\ee
and
\be
{\cal Q}^{(0)}_{\alpha_{3}\alpha_{4},\alpha_{1},\alpha_{2}}(1)
=-\tilde{{\cal V}}_{1}(\alpha_{3},\alpha_{4};\alpha_{1},\alpha_{2};1)
+\tilde{{\cal V}}_{1}(\alpha_{1},\alpha_{3};\alpha_{4},\alpha_{2};1)
+\tilde{{\cal V}_{1}}(\alpha_{1},\alpha_{4};\alpha_{3},\alpha_{2};1)
\nonumber
\ee
\be
S^{(2)}(1)=\sum_{i}<\nabla_{i}m_{\alpha}(1)\nabla_{i}m_{\alpha}(1)>
\nonumber
\ee
where there is no summation over $\alpha$.  These results give a 
closed solution for the two-point correlation functions at
zeroth-order.  This analysis of this lowest-order solution is
given in the next section.  In the following section we analyze
the four-point correlation functions needed in order to extract
the two-point correlation functions at second order.

\section{Zeroth Order Theory For Two Point Correlation Functions}

The equations of motion at zeroth order for the two-point correlation
functions are given by the coupled set of Eqs.(\ref{eq:133}) and 
(\ref{eq:134}) with the 
lowest order contributions
for $Q_{2}$ and $\tilde{Q}_{2}$ given the leading order terms
in Eqs.(\ref{eq:169}) and (\ref{eq:170}). Inserting
\be
Q_{2}^{(0)}(12)=\Omega (1\bar{1})G(\bar{1}2)
\nonumber
\ee
\be
\tilde{Q}_{2}^{(0)}(12)=\Omega (1\bar{1})G_{Mm}(\bar{1}2)
\nonumber
\ee
into Eqs.(\ref{eq:133}) and (\ref{eq:134}) and explicitly writing 
the vector labels, we obtain
\be
-i\left[\tilde{\Lambda}(1)+\omega_{0}(1)\right]
G_{M_{\alpha_{1}}m_{\alpha_{2}}}^{(0)}(12)
=\delta (12)\delta_{\alpha_{1},\alpha_{2}}
\nonumber
\ee
\be
i\left[\Lambda (1)-\omega_{0}(1)\right]
G_{\alpha_{1},\alpha_{2}}^{(0)}(12)
=-\int
d\bar{1} ~\Pi_{0} (1\bar{1})G_{M_{\alpha_{1}}m_{\alpha_{2}}}^{(0)}(\bar{1}2)
 ~~~,
\nonumber
\ee
where we have defined
\be
\omega_{0} (1) =\left(g_{0}(1)+g_{1}(1)S^{(2)}(1)\right)
{\cal V}_{0}(\alpha_{1},\alpha_{1};1) ~~~.
\nonumber
\ee
We see at once that the solution to this set of equations
is diagonal in the vector-indices,
\be
G_{\alpha_{1},\alpha_{2}}^{(0)}(12)=
\delta_{\alpha_{1},\alpha_{2}}G^{(0)}(12)
\nonumber
\ee
and
\be 
G_{M_{\alpha_{1}}m_{\alpha_{2}}}^{(0)}(12)= 
\delta_{\alpha_{1},\alpha_{2}}G_{Mm}^{(0)}(12)  ~~~,
\nonumber
\ee
where $G^{(0)}(12)$ and $G_{Mm}^{(0)}(12)$ are the same quantities
found in the scalar case in I.  We summarize briefly the results
since they are needed here.  
At long times we can write
\be
\omega_{0} (t)=\frac{\omega}{t_{c}+t}
\label{eq:176}
\ee
where $\omega$ is a constant we will determine and $t_{c}$ is a
short-time cut-off which depends on details of the early-time
evolution.  One has then that
the response function is given by
\be
G_{Mm}^{(0)}(r,t_{1}t_{2})=
G_{Mm}^{(0)}(r,t_{2}t_{1})=
-i\theta (t_{2}-t_{1})
\left(\frac{t_{1}+t_{c}}{t_{2}+t_{c}}\right)^{\omega}
\frac{e^{-\frac{r^{2}}{4(t_{2}-t_{1})}}}{[4\pi (t_{2}-t_{1})]^{d/2}}
~~~,
\nonumber
\ee
where $\vec{r}=\vec{r}_{1}-\vec{r}_{2}$,
the correlation function is given by
\be
G^{(0)}(r,t_{1}t_{2})=g(0)
\left(\frac{t_{1}+t_{c}}{t_{c}}\right)^{\omega}
\left(\frac{t_{2}+t_{c}}{t_{c}}\right)^{\omega}
\frac{e^{-r^{2}/8T}}{(8\pi T)^{d/2}}~~~,
\label{eq:157}
\ee
where $g(0)$ is the on-site value of the initial correlation
function, and it is convenient to define
\be
T=\frac{t_{1}+t_{2}}{2} ~~~.
\nonumber
\ee

If we are to have a self-consistent scaling equation then the 
autocorrelation function $(r=0)$, at large equal times 
$t_{1}=t_{2}=t$, must
satisfy
\be
S^{(0)}(t)=g(0)\left(\frac{t}{t_{c}}\right)^{2\omega}
\frac{1}{(8\pi t)^{d/2}}
\equiv A_{0}t ~~~.
\nonumber
\ee
Clearly this result fixes the exponent
\be
\omega=\frac{1}{2}(1+\frac{d}{2})
\nonumber
\ee
and the amplitude
\be
A_{0}=\frac{1}{(t_{c})^{2\omega}}
\frac{g(0)}{(8\pi )^{d/2}}
~~~.
\label{eq:184}
\ee

Eq.(\ref{eq:157}) 
can be rewritten in the convenient form
\be
G^{(0)}(r,t_{1}t_{2})=\sqrt{S^{(0)}(t_{1})S^{(0)}(t_{2})}
\Phi_{(0)}(t_{1},t_{2})
e^{-\frac{1}{2}r^{2}/(4T)}
\nonumber
\ee
where 
\be
\Phi_{(0)}(t_{1},t_{2})=\left(\frac{\sqrt{t_{1}
t_{2}}}
{ T}
\right)^{d/2}~~~.
\nonumber
\ee
The nonequilibrium exponent is defined in the long-time limit by
\be
\frac{G^{(0)}(0,t_{1},t_{2})}
{\sqrt{S^{0}(t_{1})S^{0}(t_{2})}}=\Phi_{(0)}(t_{1},t_{2})
=\left(\frac{\sqrt{t_{1}t_{2}}}
{T}\right)^{\lambda}
\nonumber
\ee
and we obtain the OJK result
$\lambda =\frac{d}{2}$.
Looking at equal times we have the auxiliary field scaling function
\be
f_{0}(x)=\frac{G^{(0)}(r,tt)}{S^{(0)}(t)}=e^{-x^{2}/2} ~~~,
\nonumber
\ee
where the scaled length is defined by
$\vec{x}=\vec{r}/L(t)$, and 
the growth law is given by
$L^{2}(t)=4t$.
The exponent $\nu$, defined by Eq.(\ref{eq:8}), is zero in the OJK
approximation.

\subsection{Four-Point Correlation Functions at First Order}

If we are to evaluate $G^{(2)}$ and $G_{Mm}$ to second order, we see that we
must evaluate the four-point quantities $G^{(4)}$ and $G_{Mmmm}$ to first 
order in the vertex expansion.  This requires the evaluation of 
$Q_{4}$ and $\tilde{Q}_{4}$.
Using the same techniques developed in evaluating $Q_{2}$ and
$\tilde{Q}_{2}$ we find
\be
\tilde{Q}_{4}(1234)=\Omega (1\bar{1})G_{Mmmm}(\bar{1}234)
+\tilde{S}_{1}(1\bar{2}\bar{3}\bar{4})P_{M}(\bar{2}\bar{3}\bar{4},234)
\label{eq:169a}
\ee
where $\Omega (12)$ and $\tilde{S}_{1}(1234)$ are defined by
Eqs.(\ref{eq:172})
and (\ref{eq:174}) respectively, while
\be
P_{M}(2'3'4',234)=G_{Mm}(4'2)\left[G(3'3)G(2'4)+G(3'4)G(2'3)\right]
\nonumber
\ee
\be
+G_{Mm}(4'3)\left[G(3'2)G(2'4)+G(3'4)G(2'2)\right]
\nonumber
\ee
\be
+G_{Mm}(4'2)\left[G(3'2)G(2'3)+G(3'3)G(2'2)\right] ~~~.
\label{eq:204}
\ee
We also have
\be
Q_{4}(1234)=\Omega (1\bar{1})G^{(4)}(\bar{1}234)
+\tilde{S}(1\bar{2}\bar{3}\bar{4})P(\bar{2}\bar{3}\bar{4},234)
\label{eq:171}
\ee
where $\tilde{S}(1234)$ is given by Eq.(\ref{eq:173}) and
\be
P(2'3'4',234)=G(4'2)\left[G(3'3)G(2'4)+G(3'4)G(2'3)\right]
\nonumber
\ee
\be
+G(4'3)\left[G(3'2)G(2'4)+G(3'4)G(2'2)\right]
\nonumber
\ee
\be
+G(4'2)\left[G(3'2)G(2'3)+G(3'3)G(2'2)\right] ~~~.
\nonumber
\ee

Inserting Eq.(\ref{eq:169a}) for $\tilde{Q}_{4}$ into Eq.(\ref{eq:125}) 
we see that we
can do a partial integration and write
\be
G_{Mmmm}(1234)=G_{Mm}^{(0)}(1\bar{1})
i\tilde{S}_{1}(\bar{1}\bar{2}\bar{3}\bar{4})P_{M}(\bar{2}\bar{3}\bar{4},234)
\label{eq:173a}
\ee
where $\tilde{S}_{1}$ is defined by Eq.(\ref{eq:173a}).  Using the 
symmetry properties
of $P_{M}(\bar{2}\bar{3}\bar{4},234)$ we can show that Eq.(\ref{eq:173a})
can be
written as
\be 
G_{Mmmm}(1234)=G_{Mm}^{(0)}(1\bar{1})
V_{s}(\bar{2};\bar{1}\bar{3}\bar{4})P_{M}(\bar{2}\bar{3}\bar{4},234)
\label{eq:174a}
\ee
where
\be
V_{s}(2;134)=\frac{i}{2}\tilde{V}_{0}(1234)
+i\tilde{V}_{L,s}(2,134)+i\tilde{V}_{1,s}(2,134)
\nonumber
\ee
and the symmetrized vertices are given by
\be
\tilde{V}_{L,s}(2,134)=\tilde{V}_{L}(2134)+\tilde{V}_{L}(2,413)
+\tilde{V}_{L}(2,314)
\nonumber
\ee
and
\be 
\tilde{V}_{1,s}(2,134)=\tilde{V}_{1}(2134)+\tilde{V}_{1}(2,413) 
+\tilde{V}_{1}(2,314) ~~~. 
\nonumber
\ee
In turn Eqs.(\ref{eq:174a}) and (\ref{eq:171}) can be put back into 
Eq.(\ref{eq:92}) with $n=4$ to obtain
$G^{(4)}$. After manipulations it can be written in the
properly symmetric form
\be
G^{(4)}(1234)=\frac{1}{3}G_{mM}^{(0)}(1\bar{1})
V_{s}(\bar{1};\bar{2}\bar{3}\bar{4})P(\bar{2}\bar{3}\bar{4},234)
+\frac{1}{3}G_{mM}^{(0)}(2\bar{1}) 
V_{s}(\bar{1};\bar{2}\bar{3}\bar{4})P(\bar{2}\bar{3}\bar{4},134)
\nonumber 
\ee 
\be 
+\frac{1}{3}G_{mM}^{(0)}(3\bar{1})  
V_{s}(\bar{1};\bar{2}\bar{3}\bar{4})P(\bar{2}\bar{3}\bar{4},124)
+\frac{1}{3}G_{mM}^{(0)}(4\bar{1})  
V_{s}(\bar{1};\bar{2}\bar{3}\bar{4})P(\bar{2}\bar{3}\bar{4},123)
~~~.
\nonumber
\ee
We see after these manipulations that all of the first order
corrections have been combined into a single
vertex.

\section{Two-Point Correlation Function at Second Order}

\subsection{General Equations}

Given $G^{(4)}$ and $G_{Mmmm}$ at first order, we can return to
Eqs.(\ref{eq:169a}) and (\ref{eq:170}) to obtain $Q_{2}$
and $\tilde{Q}_{2}$ to second order.  These in turn are put back into
Eqs.(\ref{eq:133}) and (\ref{eq:134}) to obtain the second-order
results for the two-point correlation functions.  We focus here on 
the correlation function.  After a
single integration of Eq.(\ref{eq:134}) we have
\be
G(12)=G^{(0)}(12)
+G_{mM}^{(0)}(1\bar{1})\frac{1}{3}V_{s}(\bar{1};\bar{2}\bar{3}\bar{4})
G_{4}(\bar{2}\bar{3}\bar{4}2)
+G^{(0)}(1\bar{1})V_{s}(\bar{2};\bar{1}\bar{3}\bar{4})
G_{Mmmm}(\bar{2}\bar{3}\bar{4}2)
\nonumber
\ee
\be
=G^{(0)}(12)
+G_{mM}^{(0)}(1\bar{1})\frac{1}{3}V_{s}(\bar{1};\bar{2}\bar{3}\bar{4})
\frac{1}{3}V_{s}(\bar{1}';\bar{2}'\bar{3}'\bar{4}')
\nonumber
\ee
\be
\times\Biggl[G_{mM}^{(0)}(2\bar{1}')
P(\bar{2}'\bar{3}'\bar{4}',\bar{2}\bar{3}\bar{4})
+G_{mM}^{(0)}(\bar{2}\bar{1}')
P(\bar{2}'\bar{3}'\bar{4}',2\bar{3}\bar{4})
+G_{mM}^{(0)}(\bar{3}\bar{1}')
P(\bar{2}'\bar{3}'\bar{4}',2\bar{2}\bar{4})
+G_{mM}^{(0)}(\bar{4}\bar{1}')    
P(\bar{2}'\bar{3}'\bar{4}',2\bar{2}\bar{3})\Biggr]
\nonumber
\ee
\be
+G^{(0)}(1\bar{1})V_{s}(\bar{2};\bar{1}\bar{3}\bar{4})
G_{Mm}^{(0)}(\bar{2}\bar{1}')V_{s}(\bar{2}';\bar{1}'\bar{3}'\bar{4}')
P_{M}(\bar{2}'\bar{3}'\bar{4}',2\bar{3}\bar{4})
~~~.
\nonumber
\ee
This last term simplifies since, because of causality, only the
term proportional to $G_{Mm}(\bar{2}'2)$ in 
$P_{M}(\bar{2}'\bar{3}'\bar{4}',2\bar{3}\bar{4})$ survives, and
\be
V_{s}(\bar{1}';\bar{2}'\bar{3}'\bar{4}')
P(\bar{2}'\bar{3}'\bar{4}',\bar{2}\bar{3}\bar{4})
=6V_{s}(\bar{1}';\bar{2}'\bar{3}'\bar{4}')
G(\bar{2}'\bar{2})G(\bar{3}'\bar{3})G(\bar{4}'\bar{4})
~~~.
\nonumber
\ee
We then have the final formal expressions:
\be
G(12)=G^{(0)}(12)+G^{(S)}(12)+G^{(U)}(12)+G^{(U)}(21)
\nonumber
\ee
where the {\it symmetric} contributions are given by
\be
G^{(S)}(12)=\frac{2}{3}G_{mM}^{(0)}(1\bar{1})G_{mM}^{(0)}(2\bar{1}')
V_{s}(\bar{1};\bar{2}\bar{3}\bar{4})
V_{s}(\bar{1}';\bar{2}'\bar{3}'\bar{4}')
G^{(0)}(\bar{2}'\bar{2})G^{(0)}(\bar{3}'\bar{3})G^{(0)}(\bar{4}'\bar{4})
\nonumber
\ee
and the {\it unsymmetric} contributions are given by
\be
G^{(U)}(12)=
2G_{mM}^{(0)}(1\bar{1})G^{(0)}(2\bar{2}')
V_{s}(\bar{1};\bar{2}\bar{3}\bar{4})
V_{s}(\bar{1}';\bar{2}'\bar{3}'\bar{4}')
G_{mM}^{(0)}(\bar{2}\bar{1}')G^{(0)}(\bar{3}'\bar{3})
G^{(0)}(\bar{4}'\bar{4})
~~~.
\nonumber
\ee

The detailed analysis of these contributions to the correlation 
function follows
closely the analysis developed in detail in I.  
Indeed if the full vertex $V_{s}$ is replaced by $(i/2)\tilde{V}_{0}$
and $n$ set to $1$, these equations reduce to those found in I.
One can again carry
out explicitly the internal spatial integrations.
Among the new elements in this analysis is the treatment of
the gradient insertions  in the vertices, 
and the internal vector sums.
One must also introduce the parameter,
\be
g=g_{1}(1){\cal V}_{0}(\alpha_{1};\alpha_{1}:1)S(1)
\nonumber
\ee
which,
as anticipated earlier, requires that $g_{1}(1)$ go as $L^{-1}$ for long
times with an amplitude which is determined as part of the scaling
structure.

\subsection{Extraction of Indices}

As in I, all of the various log singularities found in
second order, and arising from internal time integrations, can be absorbed
into expressions for the exponents $\omega$, $\lambda$ and $\nu$.

At second 
order in the vertex expansion the exponents are determined by the set
of equations:
\be
\lambda =\frac{d}{2}+\omega^{2}\frac{2^{d+1}}{3^{d/2}}H_{S}
\label{eq:10}
\ee
and
\be
\frac{\nu}{2}=\omega^{2}2^{d+1}\left[H_{U}+\frac{H_{S}}{3^{d/2}}\right]
\label{eq:11a}
\ee
where the quantities $H_{U}$ and $H_{S}$ are given below.
The condition that the growth law $L\approx t^{1/2}$ be maintained order
by order in perturbation theory determines the parameter $\omega$, and,
as in I,  can be expressed in terms of
the exponent $\nu$:
\be
2\omega +\frac{\nu}{2}=1+d/2 ~~~.
\label{eq:12a}
\ee
There are two intrinsically different contributions, $H_{U}$ and $H_{S}$,
 to the second-order
expressions for the correlation functions which come from graphs with
different structures.  
$H_{U}$ and $H_{S}$, which depend only on the parameters $\omega$, $g$,
$d$ and $n$, are defined in 
terms of a set
of auxiliary quantities:
\be
H_{S}=Q_{S}^{(0)}M_{d}+Q_{S}^{(2)}M_{d+2}
\nonumber
\ee
\be
H_{U}=Q_{U}^{(0)}K_{d}+Q_{U}^{(6)}K_{d}^{(6)}+Q_{U}^{(9)}K_{d}^{(9)}
\nonumber
\ee
where
\be
Q_{S}^{(0)}=\frac{1}{2(n+2)}\left(1-\frac{d}{\omega}S^{(2)}\right)
\nonumber
\ee
\be
Q_{S}^{(2)}=\frac{d}{18\omega^{2}}\left[(d+5)S^{(3)}+2(d-1)S^{(4)}\right]
\nonumber
\ee
\be
Q_{U}^{(0)}=\frac{3}{2(n+2)}\left(1-\frac{d}{\omega}S^{(2)}\right)
\nonumber
\ee
\be
Q_{U}^{(6)}=\frac{2}{\omega^{2}}S^{(3)}
\nonumber
\ee
\be
Q_{U}^{(9)}=\frac{4}{\omega^{2}}S^{(4)}
\nonumber
\ee
and the $S^{(i)}$ are basically the result of internal $n$-vector sums:
\be
S^{(2)}=-\frac{2(n-1)}{3n}+\frac{(n+2)}{3}g
\nonumber
\ee
\be
S^{(3)}=\frac{2(n-1)}{n(n+2)}+n g^{2}
\nonumber
\ee
\be
S^{(4)}=\frac{(n-2)(n-1)}{n^{2}(n+2)}-\frac{2(n-1)}{n}g+g^{2}
\nonumber
~~~.
\ee
The constants $g$ and  $\omega$, parameterize the scaling 
properties of the nonlinear terms in the equation of motion for the 
auxiliary field. 
Finally we have 
the $d$-dependent integrals
\be
M_{d}=\int_{0}^{1}dz \frac{z^{d/2-1}}{[1+z]^{d}}=
\frac{1}{2}\frac{\Gamma^{2}(d/2)}{\Gamma (d)} ~~~,
\label{eq:8}
\ee
\be
K_{d}^{(0)}=\int_{0}^{1}dz \frac{z^{d/2-1}}{[(1+z)(3-z)]^{d/2}}
~~~,
\label{eq:7}
\ee
\be
K_{d}^{(6)}=\frac{d}{4}\int_{0}^{1}dz \frac{z^{d/2-1}}{[(1+z)(3-z)]^{d/2}}
\frac{z}{(1+z)^{2}}\left[1-\frac{2(1-z)}{(3-z)}+(d+2)\frac{(1-z)^{2}}
{(3-z)^{2}}\right]
\nonumber
\ee
\be
K_{d}^{(9)}=-\frac{d}{4}\int_{0}^{1}dz \frac{z^{d/2-1}}{[(1+z)(3-z)]^{d/2}}
\frac{z}{(1+z)(3-z)}\left[1-(d+2)\frac{(1-z)}{(3-z)}\right] ~~~.
\label{eq:7a}
\ee
If we set $n=1$ and $g=0$  then this set of
equations reduces to that found in I.  Then Eqs.(\ref{eq:11a}) and
(\ref{eq:12a}) can be solved for $\omega$ and $\nu$ and the results
inserted in Eq.(\ref{eq:10}) to obtain the index $\lambda$ as given
in I.

The parameters $\omega$ and $g$ should be thought of as being
generated  by some type of 
renormalization group analysis. Carrying forward the RG analogy,
these parameters are to be determined as part of finding the scaling
fixed point in the problem.  This process is similar to 
finding a fixed point Hamiltonian in critical phenomena.
The parameter $\omega$ occurs naturally at lowest order in the perturbation
theory expansion, while $g$ naturally arises at 
second order in the expansion.  $\omega$ and $g$ are determined by the 
requirements that the scaling law $L\approx t^{1/2}$ and the index relation
$\nu =2\lambda -d$ hold at all orders.  The maintenance of the growth law
leads to the condition given by Eq.(\ref{eq:12a}).  The requirement
$\nu =2\lambda -d$ is enforced by choosing
\be
H_{U}=0 ~~~.
\label{eq:25a}
\ee  

While there are many possible ways of extracting explicit numbers for
the indices from the perturbation expansion just described, 
we discuss two here.
In the {\it expansion} method we set
$\omega =\omega_{0}=\frac{1}{2}(1+\frac{d}{2})$
in the second-order terms and obtain the indices directly.  In the 
second method we look for a self-consistent solution of 
Eqs.(\ref{eq:12a}) and (\ref{eq:25a}) for $g$ and
$\omega$. For large $d$ and $n$ these two approaches are
equivalent.
While the various $d$-dependent integrals, $K_{d}^{(i)}$, etc.
can be worked out analytically for specific values of $d$,
the expressions are not very illuminating.  Numerical values for
$\lambda$, $\nu$, and $g$ are given in Tables $I$, $II$,and $III$.
Except for the values
of $\lambda$ maked by an $*$, whose significance is discussed
below, the values of $\nu$ are given by
$\nu =2\lambda -d$ and $\omega$ is given by Eq.(\ref{eq:25a}).
One sees that the self-consistent values for $\lambda$ are all close to
the OJK values.  The perturbation theory results can lead to much
larger corrections.

It is instructive to work out the large-$d$ limit analytically.  For 
general $n$, 
one finds a solution of Eq.(\ref{eq:25a}) in the limit with
\be
g=\frac{3}{(n+2)}\left(\frac{1}{4}+\frac{2(n-1)}{3n}\right)
~~~.
\nonumber
\ee
This expression for $g$
has a minimum value of $1/4$ for $n=1$, a maximum of $7/16$ for $n=2$
and then a slow decay to zero as a function of $n$.  For the scalar case
the contribution to $\nu$ and  $\lambda =d/2+\nu /2$, are given to
leading order by
\be
\nu =\frac{\sqrt{2\pi }}{96}\frac{d^{3/2}}{3^{d/2}} ~~~
\nonumber
\ee
which gives exponential decay to zero for large $d$.

For large $n$ the exponents are also given by the OJK result, with
corrections of the form
$\lambda =d/2+\lambda_{1}(d)/n+\cdots $, where the precise dependence
of $\lambda_{1}$ on $d$ is complicated.

This procedure for fixing the coefficients $\omega$
and $g$ works straightforwardly for the scalar case and generally for
$d > n$.  However, for $d < n$
one finds, for small enough $d$, that the new spin-wave contributions
(proportional to $(n-1)$ in the $S^{(i)}$) lead to a breakdown in this
process.  Solutions to Eq.(\ref{eq:25a}) do not exist and one can not 
enforce the relation $\nu =2\lambda -d$.  
In this case we have chosen
$g$ such that $H_{U}$ is a minimum. The structure of the theory for
$n > d$ needs further work.  This is just the regime where one does not 
generate stable topological defects.

\begin{table}
\begin{tabular}{|c|c|c|c|c|}
\hline
& & & & \\
$dimension$ &  &$n=1 $ & $ n=2  $&  $n=3 $ \\
& & & & \\
\hline
& $per$      & $0.5819$ &$0.8120$  & 1.1172*  \\
&$sel$      & $0.5154$ &$0.5590$  & 0.6206*  \\
$1 $&$TUG$      & $1.0$    &$0.699$   & 0.622  \\
&$OJK$      & $0.5$    &$0.5$   & 0.5  \\
&$num$      &     &   & $0.648^{b}$  \\
\hline
&$per$      & $1.0530$ &$1.2045$  & 1.5326*   \\
&$sel$      & $1.0059$ &$1.0227$  & 1.0597*   \\
$2 $&$TUG$      & $1.2887$ &$1.171$   & 1.117   \\
&$OJK$      & $1.0$    &$1.0$     & 1.0  \\
&$num$      & $1.246\pm0.02^{a}$    &     &   \\
\hline
&$per$      & $1.5375$ &$1.6284$  & 1.8240  \\
&$sel$      & $1.5024$ &$1.5082$  & 1.5216 \\
$3 $&$TUG$      & $1.6726$ &$1.618$   & 1.587  \\
&$OJK$      & $1.5$    &$1.5$     & 1.5  \\
&$num$      & $1.838\pm0.2^{a}$    &     &   \\
\hline
$large $& & $ d/2 $  &$d/2$     & d/2     \\
 & & &  &\\
\hline
\end{tabular}
\caption{Values of exponent $\lambda$.  In second column
per refers to values from 
the current theory fully  expanded, sel
refers to a self-consistent solutions from the current theory,
TUG refers to values from Ref.(\protect\onlinecite{TUG}),
OJK refers to the values from
Ref.(\protect\onlinecite{OJK}),  and num refers numerically
determined values:
a is from Ref.(\protect\onlinecite{NEQ}),
b is from Ref.(\protect\onlinecite{NBM}).
An asterisk indicates that no solution to Eq.(\ref{eq:25a})
was found}
\end{table}

\begin{table}
\begin{tabular}{|c|c|c|c|}
\hline
& & & \\
$dimension$ &  $n=1 $ & $ n=2  $&  $n=3 $ \\
& & & \\
$1$      & $7.6423632242e-01$ &$1.2448881672e+00$  & $1.3940000000e+00$*  \\
$2$      & $5.1175087572e-01$ &$9.3442756917e-01$  & $1.2310000000e+00$*   \\
$3$      & $4.2674144077e-01$ &$7.9300891113e-01$  & $1.0569791347e+00$ \\
$large $ & $0.25$ &$0.4375$  & $0.416666$ \\
 & & &  \\        
\hline
\end{tabular}
\caption{Values of parameter g}
\end{table}

\begin{table}
\begin{tabular}{|c|c|c|c|}
\hline
& & & \\
$dimension$ &  $n=1 $ & $ n=2  $&  $n=3 $ \\
& & & \\
$1$      & $3.08732e-02$ &$1.181836e-01$  & $1.7438e+00$*  \\
$2$      & $1.17040e-02$ &$4.530440e-02$  & $1.3928e+00$*   \\
$3$      & $4.80000e-03$ &$1.645200e-02$  & $4.3149e-02$ \\
 & & &  \\
\hline
\end{tabular}
\caption{Values of exponent $\nu$ }
\end{table}

\section{Conclusions}

It has been shown how one can extend the method developed previously
for a scalar order parameter 
to the case of the $n$-vector model.  The approach developed here appears to
be a rather general tool for looking at field theories
where the field is growing and showing scaling behavior.
One is able to develop a systematic
expansion in the number of labels on the non-linear vertices appearing in
the problem.  This expansion leads directly to expressions for the
anomalous dimensions in the problem.   Less generally one is then
confronted with the interpretation of the perturbation theory expansion
in a particular realization of the theory.
As organized here, the self-consistent corrections to the OJK 
results for the indices $\lambda$ and $\nu$ are typically quite small and
vanish for both large $d$ and $n$.  The large $d$ convergence is tied to 
the enforcement of the equation $\nu =2\lambda -d$ relating the indices.

The transverse degrees of freedom enter
quite differently into the problem compared with the longitudinal degrees of
freedom.  The longitudinal contributions to the non-linear terms in the
equation of motion for the auxiliary field $\vec{m}$ must be determined
self-consistently in constructing the scaling properties in the problem.
The contributions to the transverse part of the equation of motion for the
auxiliary field are, because the transverse degrees of freedom are
massless, fixed by the requirement, for self-consistency, that the 
amplitude for the transverse
order parameter fluctuations be small compared to the ordered
component.  It turns out that the transverse contributions to the 
equation of motion for the auxiliary field are sufficiently strong,
for  fixed
$n >3$ and sufficiently small $d$, that we are unable to enforce 
the condition $\nu=2\lambda -d$.  This regime requires further study. 

The point of view developed here is somewhat unanticipated.  In the most
direct approach, as discussed in some detail in I, one makes the substitution
$\vec{\sigma}=\vec{\sigma}[\vec{m}]$ into the order parameter equation
of motion to obtain the equation of motion for $\vec{m}$.  The path
taken here is quite different since the equation of motion for
$\vec{m}$ is constructed self-consistently.
The surprising point is that the quantity $\Xi_{L}(\vec{m})$,
which enters the equation of motion satisfied by the
auxiliary field, is {\bf not} determined when we insert
$\psi =\vec{\sigma}+\vec{u} $ into the order parameter equation of
motion.  It is this freedom which allows us to construct the
scaling regime form for $\Xi_{L}(\vec{m})$.

%
%


\begin{thebibliography}{99}
\bibitem{special}  The scalar case is special since one goes over
to a discrete symmetry and no vector label.
This has the effect of
rendering the nature of the expansion as developed in I a bit more
obscure.
\bibitem{1d}  The marginal cases $n=d=1$ and $n=d=2$ have different growth
laws as discussed in Ref.\onlinecite{2}.
\end{thebibliography}
\end{document}